\documentclass[11pt]{article}
\usepackage{graphicx} 
\usepackage[utf8]{inputenc}
\usepackage{amsfonts,url,epsfig,float,xcolor}
\usepackage{amsmath}
\usepackage{geometry}
\usepackage{sectsty}
\usepackage[normalem]{ulem}
\usepackage{textcomp, gensymb}
\usepackage{subcaption}
\usepackage{cleveref}

\usepackage[style=authoryear-comp]{biblatex}
\sectionfont{\sffamily\bfseries\upshape\large}
\subsectionfont{\sffamily\bfseries\upshape\normalsize}
\subsubsectionfont{\sffamily\mdseries\upshape\normalsize}
\makeatletter
\renewcommand\@seccntformat[1]{\csname the#1\endcsname.\quad}
\makeatother

\usepackage[dvipsnames]{xcolor}

\makeatletter
\def\@maketitle{%
  \begin{center}%
  \let \footnote \thanks
    {\large \@title \par}%
    {\normalsize
      \begin{tabular}[t]{c}%
        \@author
      \end{tabular}\par}%
    {\small \@date}%
  \end{center}%
}
\makeatother

\title{\bf The Squealer:  Sensification of model exploration and model misfit\footnote{The first two authors did this work as visiting scholars at the Flatiron Institute.  We also thank Sam Felgran, Riley Carlin, Nidhi Ram, and Siquan Wang for helpful collaboration and the U.S. Office of Naval Research for partial support of this work. All the data and code for this article will be freely available and posted on Github.}\vspace{.1in}}

\author{Andrew Gelman\footnote{Department of Statistics and Department of Political Science, Columbia University, New York.}, Andrew H. Jaffe\footnote{Department of Physics, Blackett Laboratory, Imperial College, London.}, Eliot Carlson\footnote{Center for Computational Mathematics, Flatiron Institute, New York.}, and Philip Greengard\footnote{Center for Computational Neuroscience, Flatiron Institute, New York.}\vspace{.1in}}

\date{12 Jun 2026\vspace{-.1in}}

\begin{document}\sloppy

\maketitle

\begin{abstract}
We introduce a method for visual and auditory feedback when exploring the fit of a model to data.  Starting with a best-fit curve fit to data, the user can drag the curve to a new position and the computer will emit a squeal, becoming louder and more unpleasant as the discrepancy between curve and data increases.  We demonstrate with four examples:  a two-parameter curve fit to golf putting data, a four-parameter curve fit to dilution assays, a fit to cosmological data sensitive to the parameters of the Big Bang model, and a nonparametric Gaussian process fit to temperature readings.
\end{abstract}

\section{The method}

\subsection{Supplementing statistical graphics with sound}

Visualization is important, not just for data exploration but also for statistical modeling \parencite{Exploratory:2004,Wickham:2006,Wickham-Cook-Hofmann:2015}.  By plotting the estimated model alongside the data, we can visualize the model and also see places where it does not fit.

For the benefit of vision-impaired researchers, tools have recently been developed that engage sound, smell, touch, muscular resistance, voice dialogue, balance, and multiple senses at once \parencite{Daye2006,Bornmann2024,Delivering:2022,Vis:2023}.  Here we discuss a different use case in which sound is used to supplement a dynamic visual display.

Sound has a much lower information bandwidth than vision, but it has the advantage of not requiring attention.  An audible signal can reach us even when our eyes and hands are occupied with dynamic graphics.  In the present paper, we develop a visual dashboard and sonic feedback to facilitate the exploration of fitted models by allowing the user to perturb the model parameters, with the feedback conveying when and where the fit is getting worse.  The goal is for the user to develop a better understanding of where the model fits the data, where it doesn't, and why a better fit is not possible.

\subsection{Motivation and formulation of the problem}

When fitting a model, often we are in a situation in which the best-fit curve systematically misses some of the data, and we would like to ``grab'' the curve and drag it to a new place.  Presumably that would create new problems---if we are starting at the best-fit curve, any alternative would represent a greater aggregate discrepancy---but we would like to see where these problems are, and which particular data points are causing the problem.

We formulate the simplest version of the problem in terms of a Bayesian model, $p(\theta|y)\propto p(\theta)\prod_{i=1}^N p(y_i|x_i,\theta)$, with the curve of interest being $g(x|\theta)=\mbox{E}(y|x,\theta)$, which we assume can be computed analytically.  The data $x_i$ and $y_i$ are scalars and $\theta$ is a parameter vector. More general curve-fitting posteriors can also be accommodated in the formalism.

\begin{figure}
\includegraphics[width=\textwidth]{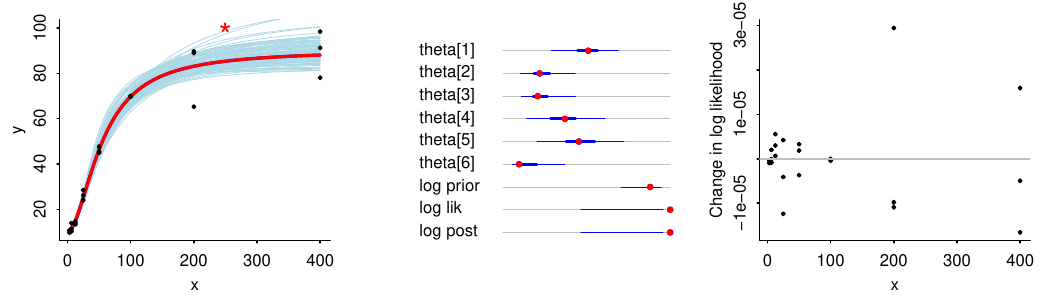}
\vspace{-.3in}
\caption{\em The basic Squealer.  {\em Left:} scatterplot of data and fitted model (dark blue curve representing the point estimate $\hat{g}$ and light blue curves representing posterior simulations $g^*$), a pseudo-data point $(x^*,y^*)$ in red, and the new curve $g^*$ in red.  {\em Center:}  dashboard showing, for each parameter $\theta_k$ in the model, the point estimate $\hat\theta_k$ as a blue dot, the posterior density from the simulations $\theta^s$ in blue, and the perturbed value $\theta_k^*$ in red.  The dashboard also shows the corresponding quantities for the log prior, likelihood, and posteriors. {\em Right:}  Relative discrepancy, $\log p(y_i|x_i,\theta^*) - \log p(y_i|x_i,\hat\theta)$, for each of the $n$ data points plotted vs.\ $x$.  In the dynamic version of the Squealer, the pseudo-data point would be specified by the user, the change in total fit, $\log p(\theta^*|x,y) - \log p(\hat\theta|x,y)$, would be conveyed by sound, and the relative discrepancies of the individual data points would be displayed as different intensities of color of the points on the graph.}
\label{fig:intro1}
\end{figure}

\begin{figure}
\includegraphics[width=\textwidth]{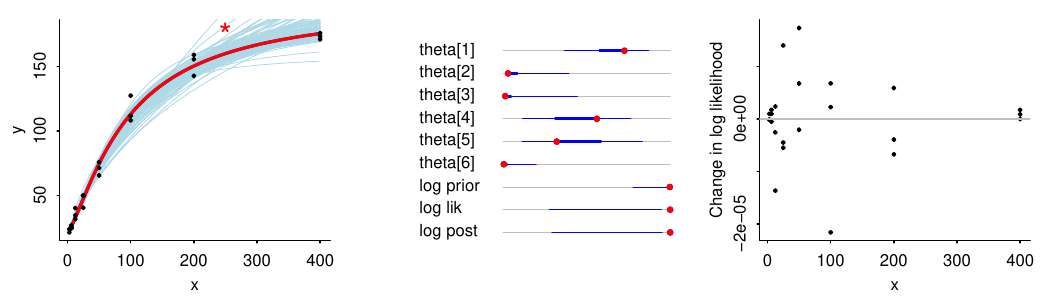}
\includegraphics[width=\textwidth]{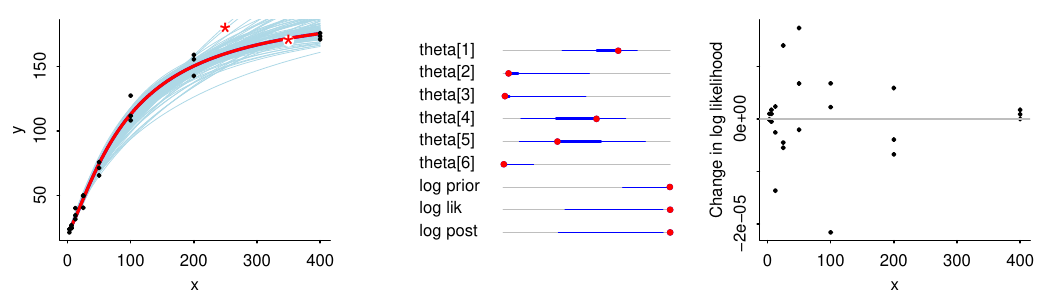}
\vspace{-.3in}
\caption{\em Example of the Squealer for data that are not consistent with the model.  In this case there is no way to get the curve close to the data: pulling up the curve to improve the fit for the point at $x=200$ degrades the fit elsewhere.  The above displays show two attempts, first adding one pseudo-data point and then adding another.  In every case, pulling toward the pseudo-data decreases the log posterior density (the lowest line on the dashboard in the middle of each row of three graphs), which makes sense given that we are starting from the posterior mode of the parameters.}
\label{fig:intro2}
\end{figure}

We assume we can fit the model and estimate $\theta$, which implies an estimate of the curve, $g(x)$, which can be overlaid on a scatterplot of $(x,y)$ and represent a starting point for our procedure.  At this point, the fit of the curve could be explored by moving the parameter vector $\theta$, which would in turn move $g(x)$.

Our key idea is to do this the other way around:  instead of changing $\theta$ and seeing what this does to $g$, we move $g$ and then adjust $\theta$ accordingly.  Altering $g$ in a certain direction will not necessarily correspond to a unique choice of $\theta$, so we will do the adjustment of $\theta$ in the context of the model to provide the best possible fit (more specifically, maximize the posterior density, $p(\theta|y)$) in a way that is consistent with $g(x|\theta)$ being perturbed in the desired direction.

Figure \ref{fig:intro1} demonstrates the idea in a setting where the data were generated from the class of models being fit.  As an example we are using a four-parameter logistic model fit to laboratory assays, as discussed further in Figure \ref{sec:dilution}.  The user should be able to explore the fit by moving the curve on the space of $(x,y)$ and seeing how this translates to changes in the parameters $\theta$ and what this does to the fit of the model to the individual data points and in total.  A challenge here is that, if $\theta$ is high-dimensional, we do not want to require the user to specify all its parameters; rather, we want to be able to move the curve $g(x|\theta)$ in some general direction and let the parameter vector $\theta$ follow in a way that is most consistent with the data.

Figure \ref{fig:intro2} shows an example where the data were generated from a model outside the fitted family, so that the best-fit curve (in blue on the left plots) has systematic discrepancies with the data.  The two rows of plots show two attempts to get a better fit by dragging the curve toward the pseudo-data points shown in red on the left plots.

If we knew that the assumptions underlying our model were correct, there would be no reason to do this, as we could just summarize uncertainty using the Bayes posterior distribution.  Realistically, though, our models will be wrong, and we are developing a set of tools for a new and dynamic way to understand misfit to data.

The present paper focuses on the statistical method and computational concerns, not on the dynamic graphics or the sonic and visual implementation.

\subsection{Perturbing the posterior mode}

The simplest version of the Squealer works with the estimate $g(x|\hat{\theta})$, where $\hat{\theta}$ is the mode of the posterior density of the fitted model, $p(\theta|y)$.  The user can drag the curve in some direction by pulling it toward to a newly specified pseudo-data point, $(x^*,y^*)$.  We do this by multiplying the posterior density by a factor, $\mbox{normal}(y^*| g(x^*|\theta),\sigma^*)$, where $\sigma^*$ is set to a value small enough that the new point will exert a strong enough ``gravitational force'' but not so small as to induce computational instability.  A reasonable first choice is half the posterior standard deviation of $g(x^*)$, that is, we set
\begin{equation}\label{eq:sigma_star}
\sigma^* = 0.5 \,\mbox{sd}(g(x^*)|y) = 0.5 \!\int\! \mbox{sd}(g(x^*|\theta)) p(\theta|y)dy,
\end{equation}
which should be easy enough to calculate, exactly or approximately, given that the model has already been fit.  We then compute the posterior mode, $\theta^*$, for this augmented model.  Using the normal distribution as a baseline, we should expect this to shift the curve approximately 80\% (that is, $\frac{1^2}{1^2 + 0.5^2}$) toward the pseudo-data point.

In the computation, we use the existing estimate, $\hat{\theta}$, as the initial value for the optimization.  There could be computational challenge, however, if the pseudo-data point is far from the fitted curve.  One approach is to check if $|y^* - \mbox{E}(g(x^*|y))| > \mbox{sd}(g(x^*|y)$ and, if so, shift $y^*$ gradually, moving it by $\mbox{sd}(g(x^*|y)$ at a time until we eventually reach the specified value.

In assuming a normal distribution for this gravitational force, we are \emph{not} assuming that the likelihood for the data $y_i$ necessarily follow a normal distribution.  The pseudo-data point does not have to take the mathematical form of an additional measurement from the model, $p(y|x,\theta)$; rather it is entirely a mathematical and computational tool for gradually moving the curve from $\hat{g}$ to a new estimate, $g^*=g(x|\theta^*)$, obtained by computing the mode  of the augmented posterior distribution of $\theta$.  For example, if $p(y|x,\theta)$ is a discrete-data model such as an ordered logistic regression of a response with five categories, it would still make sense to choose a continuous value for $y^*$, as we are using it to shift the continuous curve $g$, not to represent a possible data point.

For a complicated model it can make sense to move the curve in more elaborate ways, by pulling it toward multiple pseudo-data points, $(x^*_i,y^*_i), \ i=1,\dots,N^*$.  The same logic holds; we merely multiply the posterior density by $\prod_{i=1}^{n^*}\,\mbox{normal}(y^*_i| g(x^*_i|\theta),\sigma^*_i)$, setting each $\sigma^*_i$ using some rule such as (\ref{eq:sigma_star}), and then obtain a new estimate, $\theta^*$, and a corresponding new curve, $g^*=g(x|\theta^*)$.

In either case, the new curve $g^*$ represents the best fit prediction given the model and data, under the soft constraint that the curve passes near the new point or points, $(x^*,y^*)$.  When a model is complicated and has many parameters, we do not want to be in the position where we have to specify the entire curve; rather, we want to be able to gently drag it and then let the data dictate the best fit given our soft constraints.  Figures \ref{fig:intro1} and \ref{fig:intro2} show examples.

On the computer, the user should be able to perform this operation in one of two ways, either by clicking on the location of the pseudo-data points or by clicking on a point or points on the fitted curve $\hat{g}$ and then using the mouse or joystick to move these points to new places, which become the pseudo-data points $(x^*,y^*)$.  In either case, this would be performed using dynamic graphics on the plot of data and fitted curve, and then in the background the computer would re-fit the model and display the updated curve.

Depending on the speed of computation, there are different ways that the updated curve could be estimated and displayed.  If the fit is fast enough, the computer could re-fit the model on the fly as the curve is being dragged so that the user could see the updated curve moving in real time.  If real-time updating is impossible, it could be possible to compute the updated model using an annealing-like procedure in which the gravitational scale, $\sigma^*$, starts at a high value and then gradually decreases to zero, and the fitting algorithm can provide real-time updates.

\subsection{Perturbing a posterior mean or other point estimate}

The mode is not always a good summary of the posterior distribution:  the best-fit curve will always overfit to some extent, and even in a simple multivariate normal posterior distribution, the mode will not be in the typical set as the dimensionality increases \parencite{Typical:2020}.  And, even in low dimensions, the posterior mode can have problems.  For a hierarchical model with weak priors the joint mode occurs at the degenerate point at which all the local parameters are equal to each other and the group-level variance is zero.

For these reasons, we often choose posterior summaries other than the mode.  Two alternatives are the mean and median, either of which can be computed using posterior simulations $\theta^s,\ s=1,\dots,S$, or from a variational approximation to $p(\theta|y)$.  For the Squealer, we are interested in $g(x|\theta)$, not $\theta$ itself, so if we have an estimate $\hat{\theta}$ (which could be the posterior mode, mean, median, or a summary of a variational inference), we can use $\hat{g}$ defined by $\hat{g}(x) = g(x|\hat{\theta})$.

Alternatively we could take the posterior expectation of the curve itself, so that $\hat{g}(x) = \mbox{E}(g(x)) = \int\! g(x|\theta) p(\theta|y)$.  This is appealing from a theoretical perspective given that we have already defined $g(x|\theta)$ as $\mbox{E}(y|x,\theta)$, so that this new $\hat{g}$ is simply the unconditional posterior expectation.  There is a computational challenge here, though, in that this averaging would need to be evaluated at a grid of values of $x$, unlike the other approach in which the point estimate $\hat\theta$ only needs to be computed once and then it can be plugged into the formula for $g(x|\hat\theta)$.

Whatever choice of distributional summary we use, the plan is to first compute it from the posterior distribution, $p(\theta|y)$, and then on the augmented posterior distribution, $p(\theta^*|y)\propto p(\theta|y)\mbox{normal}(y^*|x^*,\sigma^*)$, using the inference from $p(\theta|y)$ to initialize the computation.

If the original model has been fit using posterior simulations, one might hope to be able to shift the curve by importance reweighting of the simulation draws using the normal density corresponding to the pseudo-data.  It would then be fast to compute the mean of $\theta$ under the augmented posterior distribution or to compute $\mbox{E}(g(x|\theta))$ in parallel at a grid of values of $x$.  Unfortunately we often will be moving the curve to values that are barely supported by the original model---that is one of the reasons for developing this method in the first place, to understand why a certain alternative fit does not work for the data at hand---, in which case simple importance weighting would not work.  It could make sense to use some sort of particle filtering algorithm, but here we will assume that we will just fit the augmented model directly.

When perturbing $g(\hat\theta)$, where $\hat\theta$ is the posterior mode, the log posterior density necessarily decreases.  When perturbing a $\hat{g}$ using the posterior mean or median or other summary, there is no guarantee that the log posterior will decrease, so it is not so clear what should be ``squealing'' to indicate discomfort with the fit.  One approach might be to follow a variational framework and define discomfort as the Kullback-Leibler discrepancy of $p(\theta^*|y)$ relative to $p(\theta|y)$.  In this case, the point estimate $\hat g$ is there for the graphical display but would not be used to summarize total error.

\subsection{Perturbing a posterior distribution}\label{perturbing.posterior}

\begin{figure}
\centerline{\includegraphics[width=\textwidth]{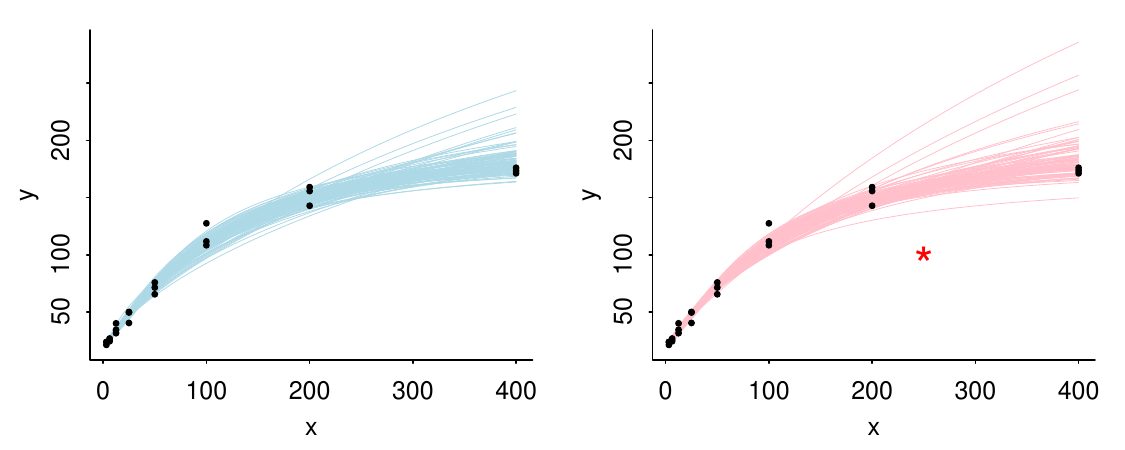}}
\vspace{-.1in}
\caption{\em Challenge of shifting the posterior distribution by pulling it toward a pseudo-data point.  {\em Left graph:} data and fitted curves $g(x|\theta^s)$ based on 100 random draws $\theta^s$ from the posterior distribution, $p(\theta|y)$, for the data and model shown in Figure \ref{fig:intro1}. {\em Right graph:}  data and pseudo-data point, along with fitted curves based on 100 random draws from the augmented posterior.  The curves on the right are constricted by the ``local gravity'' of the pseudo-data point.}
\label{fig:post1}
\end{figure}

For Bayesian applications we are often interested in the entire posterior distribution, and we would like to see what happens if it is shifted in some way.  Unfortunately, our ``gravitational'' approach leads to an artifact by which the posterior distribution of the shifted curve $g^*$ is constricted in the neighborhood of each pseudo-data point.  We would like to have a way to shift the entire distribution without pinning it in any particular place.  One idea that might seem appealing is to shift the posterior draws of $\theta$ by a fixed amount, that is, to define simulations $\theta^{*s} = \theta^s + \theta^* - \hat\theta$, but that gives poor results in examples such as this one where the dependence structure of $\theta$ is not translation invariant.  A shift on the log scale also fails.

Ultimately the challenge here is that the augmented posterior distribution contains additional information, which in some sense should not be counted when determining uncertainty in $g^*$.  For the examples in the remainder of this paper we will just shift the point estimates.

\section{Examples}

\subsection{A two-parameter model for golf putting}\label{sec:golf}

\textcite{Broadie:2018} and \textcite{Gelman:2019} presents a two-parameter model for the probability of a success of a putt in golf as a function of the distance $x$ (measured in feet) from the hole:
\begin{equation}
\label{eq:golf1}
g(x|\theta) = \left(2\Phi\left(\frac{\sin^{-1}((R-r)/x)}{\sigma_{\rm angle}}\right)-1\right)\left(\Phi\left(\frac{2}{(x+1)\,\sigma_{\rm distance}}\right)-\Phi\left(\frac{-1}{(x+1)\,\sigma_{\rm distance}}\right)\right),
\end{equation}
where $\Phi$ is the normal cumulative distribution function; $R$ and $r$ are the radius of the hole and the ball; and $\theta=(\sigma_{\rm angle}, \sigma_{\rm distance})$ are the standard deviation of the variation in the angle and the log distance of the shot, relative to the aimed angle and distance.

We fit this model to a database of 1.25 million attempted puts from pro golfers, assuming the probability of success depends only on the distance from the hole.  It would make sense to later fit a hierarchical model allowing this probability to vary by player, weather, and characteristics of the golf course, but it is reasonable to start with aggregate data to get a sense of the overall model fit.

The data binned by distance from the hole, binned into 31 categories $i$, centered at distances $x_i$ of 0.25, 1, 2, 3, \dots, with the bins becoming wider at greater distances and with the final bin cantered at 75 feet.  For each bin, we fit the model,
\begin{equation}
\label{eq:golf2}
n_i y_i\sim\mbox{binomial}(n_i, g(x_i|\theta)),
\end{equation}
where $n_i$ is the number of shots from that distance bin and $y_i$ is the proportion of those shots that went in.

\begin{figure}
\centerline{
\includegraphics[width=.7\textwidth]{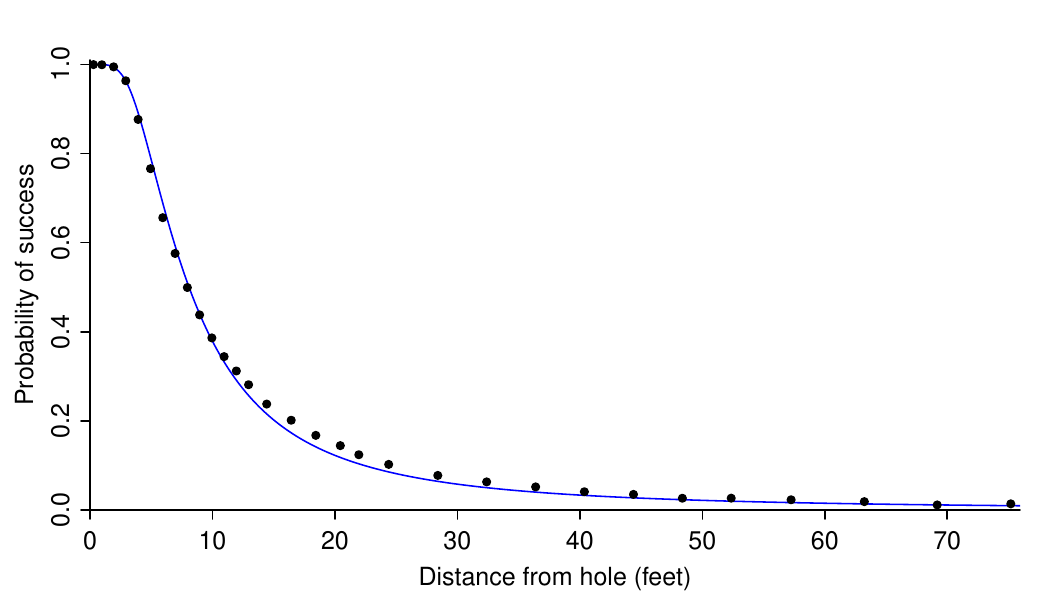}
}
\vspace{-.05in}
\caption{Data on the proportion of successful golf putts as a function of distance from the hole, along with the curve corresponding to the best-fit estimate from a specified  two-parameter model.  The model fit shows some problems for distances between 10 and 50 feet, and we would like to explore what happens if we try to move the curve toward the data.}
\label{fig:golf0}
\end{figure}

Figure \ref{fig:golf0} shows the data $(x,y)_i,\ i=1,\dots,N$, along with the best-fit curve $g(x|\hat\theta)$.  With over a million data points and only two parameters, the model is estimated so precisely that curves corresponding to posterior simulation draws, $g(x|\theta^s)$ would be visually indistinguishable from the best fit.

It is impressive how well this two-parameter curve lines up with the data---but the fit is not perfect.  There is a region between $x=10$ and $x=50$ where the curve is too low, and we would like to pull it upward.  This is a job for the Squealer!

\begin{figure}
\centerline{
\includegraphics[width=.9\textwidth]{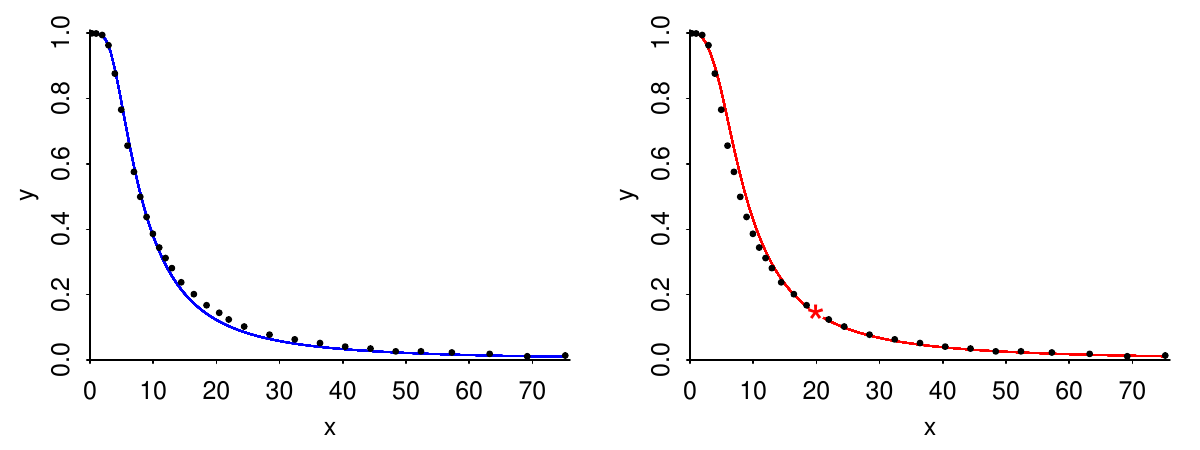}
}
\centerline{
\includegraphics[width=.9\textwidth]{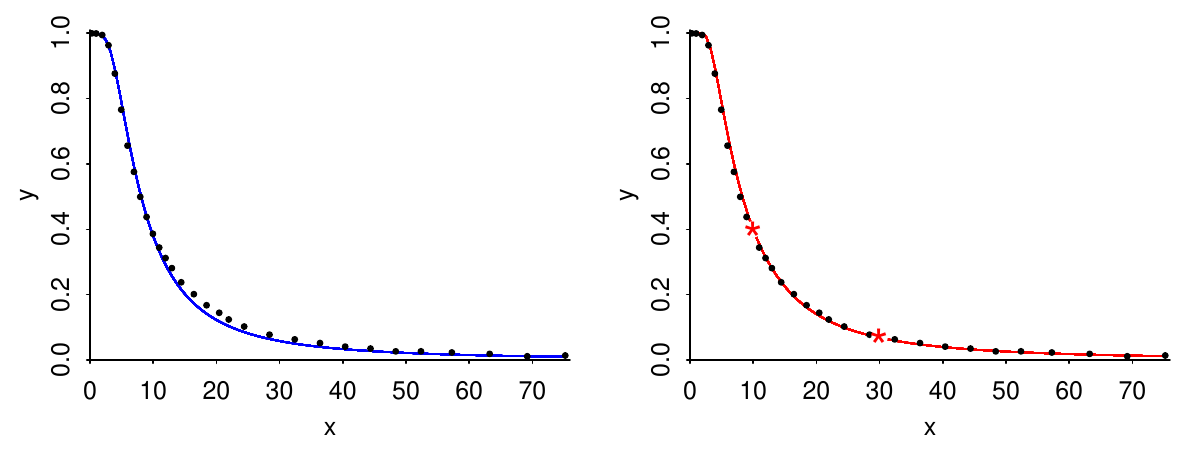}
}
\vspace{-.1in}
\caption{\em {\em Top row:}  first try to adjust the golf model to line up with the data, obtained by dragging the curve upward at one point at $x=20$.  {\em Bottom row:}  second try, dragging the curve at two points ($x=10$ and $x=30$), yields a much improved visual fit to the data.  Point-by-point details are shown in Figure \ref{fig:golf3}.  These graphs are intended to display multiple curves corresponding to posterior simulation draws of $\theta$, but in this case the posteriors are so precisely determined that no variation in the curves is visible.}
\label{fig:golf1}
\end{figure}

Because we have not yet set up the dynamic graphics, we perturb the curve by trial and error.  First we grab the value of $g(x)$ at $x=20$ and move it upward to match the data; the result is shown in the top row of Figure \ref{fig:golf1}.  This fixes the misfit between $x=20$ and $x=50$ but creates a new problem for $x$ between 5 and 15.  But we can fix this by tying the curve to two pseudo-data points, as demonstrated in the bottom row of Figure \ref{fig:golf1}.

\begin{figure}
\centerline{
\includegraphics[width=\textwidth]{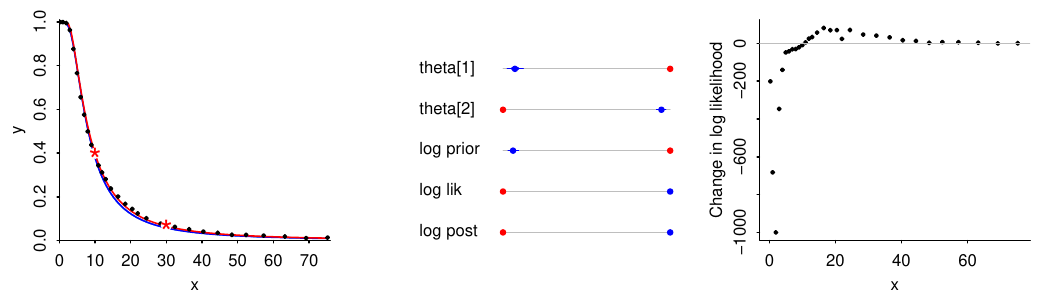}
}
\vspace{-.1in}
\caption{\em Squealer applied to the golf model.  It is possible to pull the fitted curve $g(x|\hat{\theta})$ toward two pseudo-data points and obtain a new curve $g^*=g(x|\theta^*)$ that is a better visual match to the data, but the log-posterior density for $\theta^*$ is much lower.  The rightmost plot shows that the poor fit is coming from the data points with the lowest values of $x$.}
\label{fig:golf3}
\end{figure}

We next turn to the Squealer dashboard and point-by-point summary to understand the statistical price paid by the apparently better fit. Figure \ref{fig:golf3} shows the result.  The new curve $g(x|\theta^*)$ looks good but it pays a huge price in the likelihood, coming from a few points on the left of the graph.  What is happening is that there are hundreds of thousands of putt attempts in these first few bins (consider that the bins of short putts include all the failed attempts from other distances that came close but did not go in, hence there is a pileup of data near $x=0$), hence the binomial likelihood is very strong there, so much that it does not easily tolerate even very small departures of the data $y_i$ from their expected values $g(x_i|\theta)$.

No model is perfect, and it is not ideal for the fit of the entire curve to be driven so strongly by these few data points.  \textcite{Workflow:2025} extend the model by adding an error term to account for imperfections in (\ref{eq:golf1}) and (\ref{eq:golf2}); our point here is to demonstrate how the Squealer can both facilitate a better fit to data and give insight into why the initial fitting procedure did not get there.

We include a software prototype with the golf example at \url{https://github.com/elc45/squealer-prototype}. In this interface, the starting display shows the data superimposed with the curve of the posterior mode fit. The user can click the plot to add a pseudo-data point at the location of the cursor, optionally using the sliding scale at the bottom of the console to control the ``gravitational pull'' of the point, and the curve will automatically adjust to be the mode of the new posterior augmented with the pseudo-data point. The program then makes a noise whose pitch is proportional to the decrease in log likelihood \textit{under the original model} incurred by this change in fit. Under the hood, this new pseudo-mode is found in real time using the L-BFGS optimizer. The user can subsequently add more points in different locations, adjust the gravitational pull, or remove points, with the curve responding in real time.
We include a software prototype with the golf example at \url{https://github.com/elc45/squealer-prototype}. In this interface, the starting display shows the data superimposed with the curve of the posterior mode fit. The user can click the plot to add a pseudo-data point at the location of the cursor, optionally using the sliding scale at the bottom of the console to control the ``gravitational pull'' of the point, and the curve will automatically adjust to be the mode of the new posterior augmented with the pseudo-data point. The program then makes a noise whose pitch is proportional to the decrease in log likelihood \textit{under the original model} incurred by this change in fit. Under the hood, this new pseudo-mode is found in real time using the L-BFGS optimizer. The user can subsequently add more points in different locations, adjust the gravitational pull, or remove points, with the curve responding in real time.

\subsection{A model for calibration in bioassays}\label{sec:dilution}

\begin{figure}
\centerline{
\includegraphics[width=\textwidth]{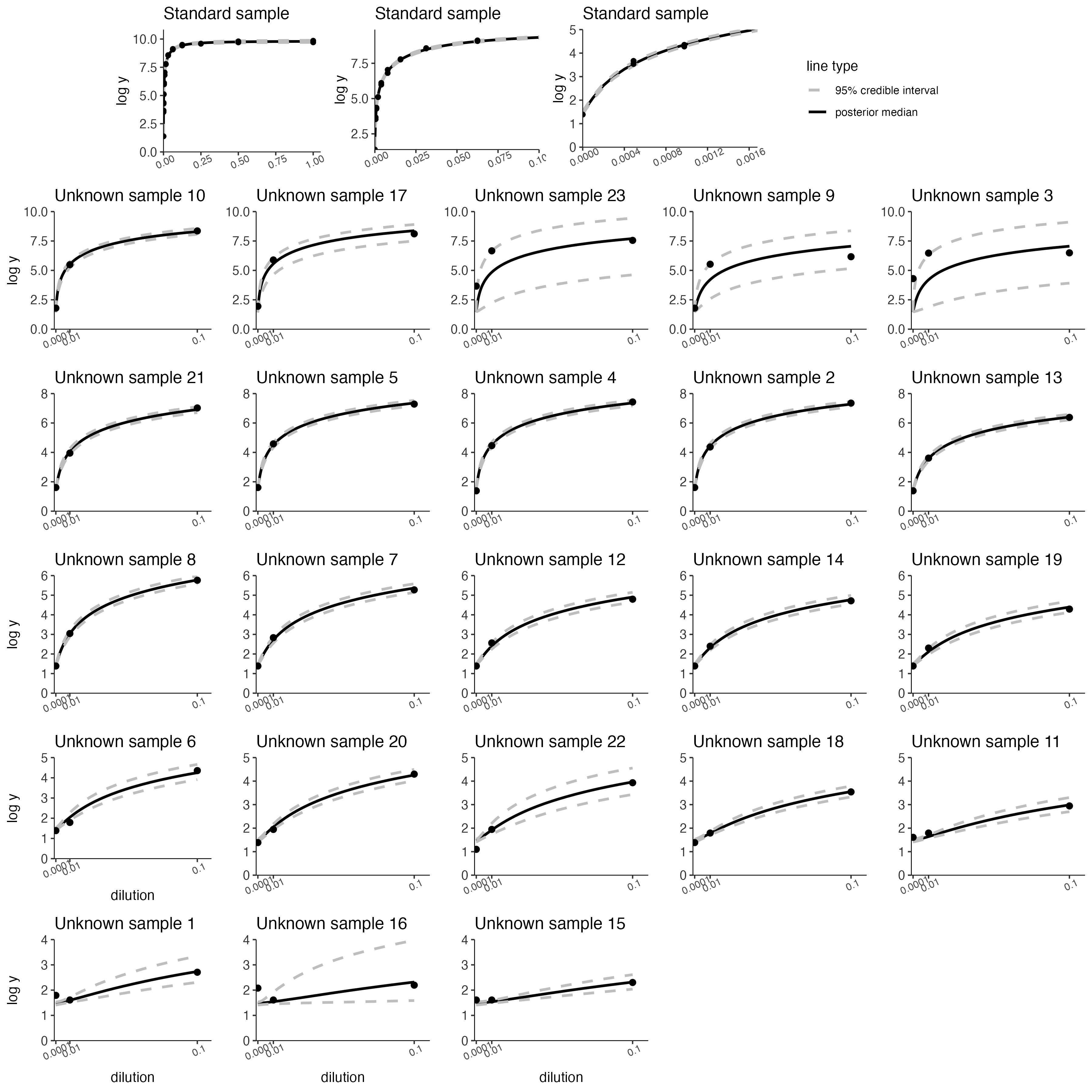}
}
\caption{\em A model for laboratory assays fit to calibration data (top row) and samples with unknown concentrations (remaining rows).  For each, data have been gathered at multiple dilutions, and the curves show expected measurement value as a function of the dilution level.  The unknown samples are displayed in decreasing order of mean measurements.  The curves for samples 23 and 3 show some misfit to the data; we explore this with the Squealer in Figure \ref{fig:dilution2}.}
\label{fig:dilution1}
\end{figure}

We next show an example of the Squealer applied to a hierarchical data structure.
\textcite{Dilution:2004} present a Bayesian approach to inference from laboratory assays in which a model is simultaneously fit to calibration data (from samples with known concentrations) and unknown samples whose concentration is then estimated.  More recently we have expanded the model by allowing for contamination.  The challenge is that sometimes a single model does not appear to fit all the data well.  Figure \ref{fig:dilution1} shows an example of a plate with several dilutions of the calibration data and 22 unknown samples.  The curves for samples 23 and 3 show some misfit to the data; we will explore this with the Squealer.

\begin{figure}
\centerline{
\includegraphics[width=\textwidth]{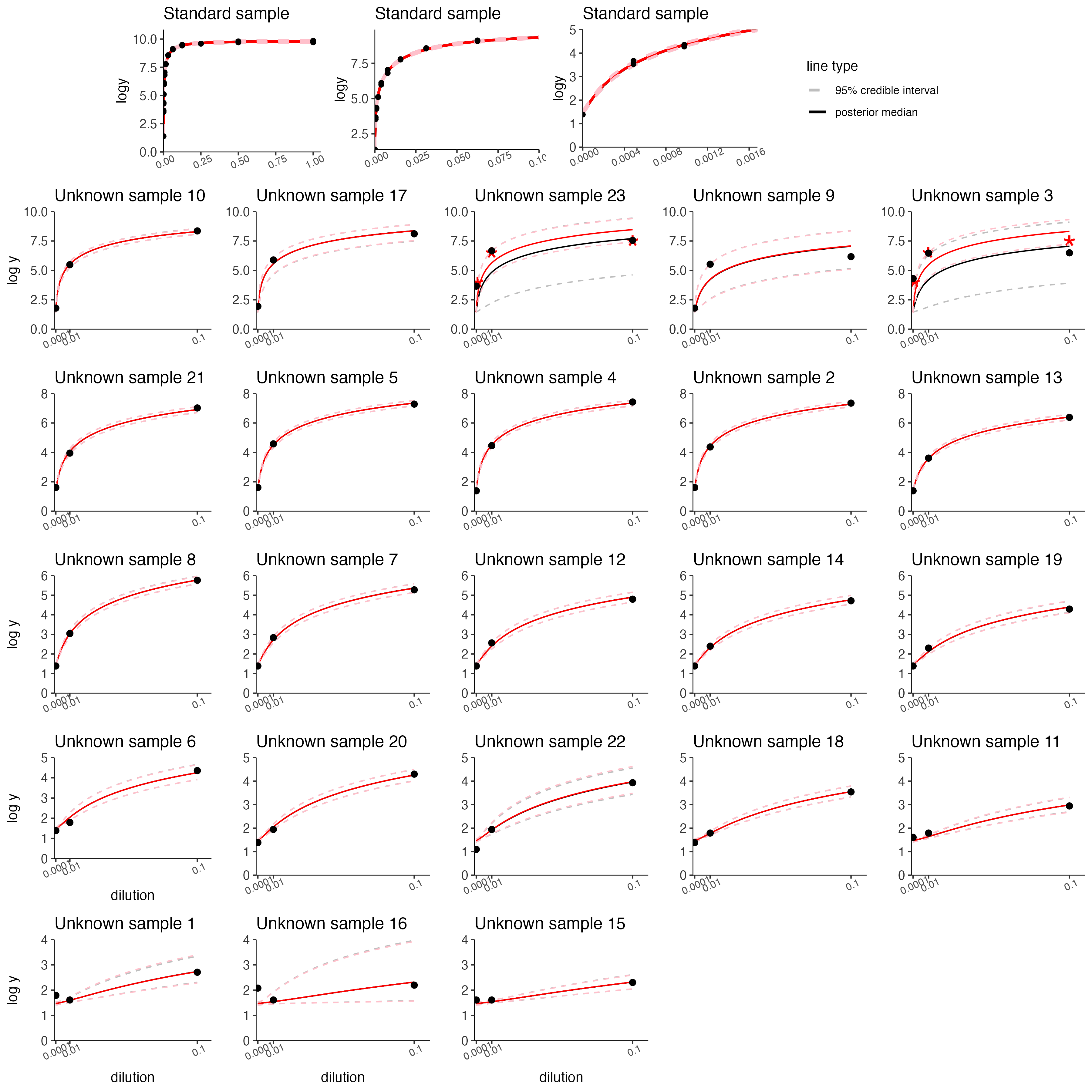}
}
\caption{\em Applying the Squealer to the data and model from Figure \ref{fig:dilution2} to perturb the curves for samples 23 and 3.  The curves fit to the original data are in black; the Squealer results are in red.  Except for the two affected samples, the two fits are essentially identical.  The next step is to understand what was constraining the original fit to be off for samples 23 and 3.}
\label{fig:dilution2}
\end{figure}

For each of samples 23 and 3 in Figure \ref{fig:dilution1}, we add three pseudo-data points (indicated by red asterisks on the plots for those samples) and re-fit the model.  Figure \ref{fig:dilution2} shows the result, with the original fit shown in black and the new fit in red.  All the parameter estimates in the model change, but the differences in the fitted curves and their uncertainties are imperceptible except for the two affected samples.

\begin{figure}
\centerline{
\includegraphics[width=\textwidth]{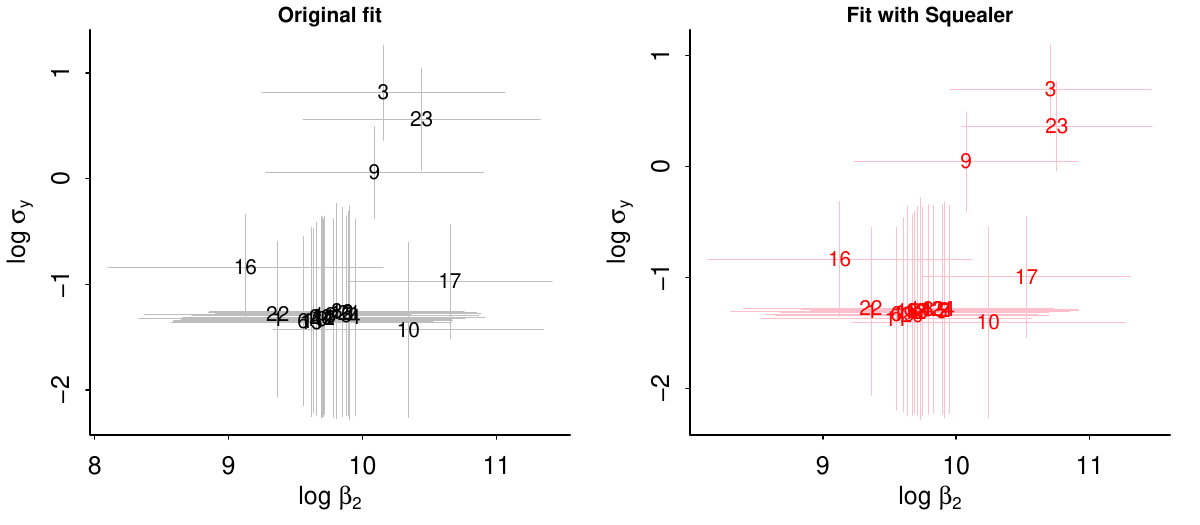}
}
\caption{\em Parameter estimates corresponding to the fitted models in Figures \ref{fig:dilution1} and \ref{fig:dilution2}.  The model has two parameters that vary across the 23 unknown samples:  $\beta_2$, which is proportional to the concentration of the compound of interest within the sample, and $\sigma_y$, the scale of modeling/measurement error.  The plots show the posterior estimates $\pm 1$ standard error for $\log \beta_2$ and $\log \sigma_y$ for each sample.  Perturbing the fit altered the inferences for samples 3 and 23 while barely changing anything for the other samples.  Note the different scales of the two graphs.}
\label{fig:dilution3}
\end{figure}

It seems, then, that it is possible to get an improved (if imperfect) fit for those two unknown samples without interfering with the overall model fit, so where is the problem?  Why did the model not fit this way originally?

The answer to this question can be seen in Figure \ref{fig:dilution3}, which displays the inferences for the parameters in the model that explain the differences between the samples.  After applying the Squealer with the pseudo-data, the estimated concentrations for samples 23 and 3 are much bigger, and the set of 23 values of the parameter $\beta_2$ no longer fit a tight lognormal distribution.  The penalty is being paid in the prior, with the cost being a higher group-level variance in the hierarchical model for the $\beta_2$'s.
Also the posterior uncertainties for those two values are much smaller; this is an unfortunate artifact of the pseudo-data implementation, as discussed in Section \ref{perturbing.posterior}.

\subsection{An example from cosmology}\label{sec:cosmology}


In the standard model of cosmology, most of the information is contained in the two-point function (power spectra or correlation functions) of astrophysical observables such as the number density of galaxies in a set of observed voxels or the intensity of the cosmic microwave background (CMB) measured at pixels on the sphere of the sky \parencite[e.g.,][]{HuSugiyamaSilk:1997,Dodelson:2003,PlanckCosmology}. These spectra or correlation functions are completely determined by the vector of underlying cosmological parameters $\theta$ (e.g., $H_0$, the Hubble constant which describes the present-day expansion rate; $\Omega_\mathrm{c}h^2$ and $\Omega_\mathrm{b}h^2$, composite quantities which describe the densities of the cold dark matter and baryons; $A_\mathrm{s}$ and $n_s$ which describe the distribution of matter in the universe at early times; $\tau$, which describes the effect of the earliest stars and galaxies on the ionization state of gas in the universe). For quantities measured on the sphere of the sky, such as the CMB, the power spectrum $C_\ell(\theta)$ determines the statistical properties of the observables, with a distribution that is typically well-described by a multivariate normal distribution. Here, $\ell$ denotes the spherical harmonic multipole, the equivalent of the Fourier mode $k$. The power spectrum is measured at spherical harmonic multipoles $2\le\ell\le\ell_\mathrm{max}$ as ${\hat C}_\ell$. The quantity $C_\ell(\theta)$ depends on the cosmological parameters in a complicated, nonlinear way. Formulated in our framework, $\ell$ is equivalent to $x$, $g(x|\theta)$ to $C_\ell(\theta)$ and $y$ to ${\hat C}_\ell$. Because of this dependence, this gives us a posterior distribution
\begin{equation}\label{eq:cmbmodel}
p(\theta|d) \propto p(\theta) p(d|\theta) = p(\theta) p\left(d|C_\ell(\theta)\right).
\end{equation}

As a slight subtlety, this posterior can be realized in several different ways. The most self-consistent (Bayesian) algorithm is to consider the multivariate normal likelihood corresponding to the model
\begin{equation}
d_p \sim \textrm{MVN}(0, \Sigma_{pp'}+N_{pp'})
\end{equation}
where the data are measured at pixels $p=1,\dots, n_\mathrm{pix}$, and $\Sigma_{pp'}$ and $N_{pp'}$ are covariance  matrices: $\Sigma_{pp'}$ depends linearly on  $C_\ell(\theta)$ and $N_{pp'}$ is a fixed noise covariance matrix, often well-approximated as diagonal. This is an accurate expression for our model of the underlying CMB field, but can be very expensive to compute, scaling as $O(n_\mathrm{pix}^3)$.

In many cases, we consider the multivariate mode of  $p(d|C_\ell)$ as estimates ${\hat C}_\ell={\hat C}_\ell(d)$, and the likelihood can then be written as $p({\hat C}_\ell|C_\ell)$. The shape around this mode can itself be approximated as a multivariate normal, although more precise approximations are sometimes available \parencite{BJK:2000}. Alternatively, the mode itself can be approximated at much lower computational cost as an unbiased estimator using the inverse spherical harmonic transform \parencite{MASTERCMB,NAMASTER}.

These developments let us treat this as a curve-fitting problem to be visualized using the Squealer. Pseudo-data enforcing
$C_{\ell^*} = {\widetilde C}_{\ell^*}$
(at some particular multipole $\ell=\ell^*$) can always be added as a new term in the likelihood
$(C_{\ell^*}-{\widetilde C}_{\ell^*})^2/(2{\bar\sigma}_{\ell^*}^2)$, no matter the form of the original likelihood. 

\begin{figure}
\centerline{
\includegraphics[width=1.0\textwidth]{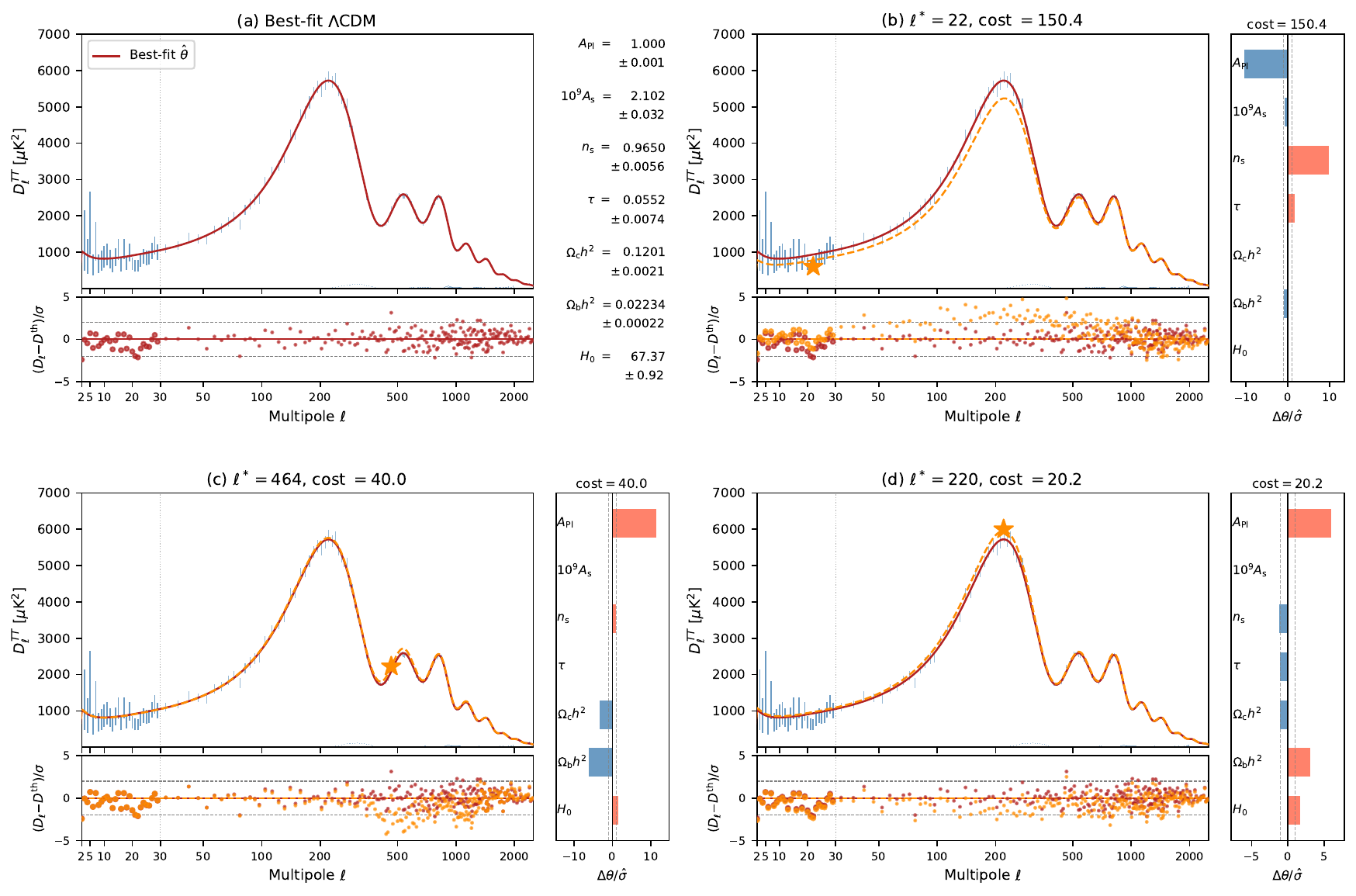}
}
\vspace{-.1in}
\caption{\em The Squealer applied to the CMB power spectrum as measured by the Planck satellite. The $x$-axis is on an idiosyncratic scale to more evenly display the data. Here, $D_\ell = \ell(\ell+1)C_\ell/(2\pi)$ as is traditionally plotted in cosmology. Panel (a) shows the data (blue error bars), the overall best-fit (posterior mode) spectrum (red curve) and residuals between the two (red points in bottom panel, standard error units), along with inferred parameter values. Panels b, c, and d show an application of the Squealer. The orange star shows the pseudo-data point, the orange curve shows the new maximum posterior, with orange points in the bottom panel showing residuals. The right sub-panel shows the shift in each of the parameters (in standard error units), and is labeled by the cost, the decrease in the log-likelihood induced by the addition of the pseudo-data.}
\label{fig:CMB1}
\end{figure}

Figure \ref{fig:CMB1}a shows measurements of the CMB power spectrum \parencite{PlanckCosmology}. For $\ell<30$, the spectrum is calculated as the mode of the full posterior (\ref{eq:cmbmodel}), whereas for $\ell\ge 30$ it is calculated by inverse spherical harmonic transform and binned in $\ell$ for display. The red curve shows the best fit model, calculated as the posterior mode.  For these calculations, we use the \texttt{plik-lite} likelihood for $\ell>30$ and the \texttt{Commander} likelihood for $\ell\le30$, in both cases only using temperature data \parencite[these codes are described in][]{PlanckLikelihood}. We model the effect of polarization data by including an informative prior on the parameter $\tau\sim \mbox{normal}(0.0543, 0.0073)$.

We can immediately see that there are regions where the residuals with respect to the curve, shown in the bottom panel, differ systematically from the data, for example the deficit of power around $\ell\simeq20$ and the positive and negative excursions around $\ell\simeq450$. What happens if we force the curve to go through these regions? In this case we care not only about the goodness of fit when new pseudo-data are added, but the induced values $\theta^*$ are physically meaningful and might have relevance for other astrophysical observations. We show values of seven parameters being fit: $A_\mathrm{Pl}$ is an instrumental nuisance parameter describing the relative amplitude of different photon frequency measurements measured by the Planck satellite; the other six are parameters describing the cosmological model \parencite{PlanckCosmology}.

The other three panels of Figure \ref{fig:CMB1} show the Squealer applied to different regions of the plot. (Unlike the other applications discussed here, we show only the new best fit, rather than a set of samples.) Figure \ref{fig:CMB1}b shows an attempt to fit to the dip at $\ell\simeq20$; there are no parameter values which modify the spectrum only in this region, and so the total cost to such a change is substantial. Figure \ref{fig:CMB1}c shows a modification at $\ell=464$, which induces less drastic changes to the spectrum and to the parameters but requires larger negative excursions to accommodate the pseudo-data. As a further example, not actually favored by the data, Figure \ref{fig:CMB1}d shows how modifying the spectrum at exactly the first peak can be accommodated at even lower cost. In the latter two cases, known physical processes account for the ability to remain a relatively good fit.

\subsection{A nonparametric Gaussian process model for temperature recordings}\label{sec:gp}
Gaussian processes (GPs) are commonly used for non-parametric modeling of time series data.
In a typical setting, a researcher or practitioner is interested in recovering an unknown function
$f$ from noisy samples from that function. A GP distribution is used as a prior
for the unknown function that enforces domain knowledge about the smoothness of the $f$.
Specifically, a simple GP model uses a normal observation model with fixed and (possibly unknown) residual standard deviation and a GP prior on the unknown function.

Here, we use this model in a simulated example, to estimate an individual's body temperature from noisy thermometer recordings. This example has the benefit that it's a commonly-used, easily interpretable model, and it's also straightforward to implement a Squealer---the pseudo-data point method is available in closed form when hyperparameters are fixed.

\begin{figure}
\centering
\begin{subfigure}[b]{0.32\textwidth}
    \centering
    \includegraphics[width=\textwidth]{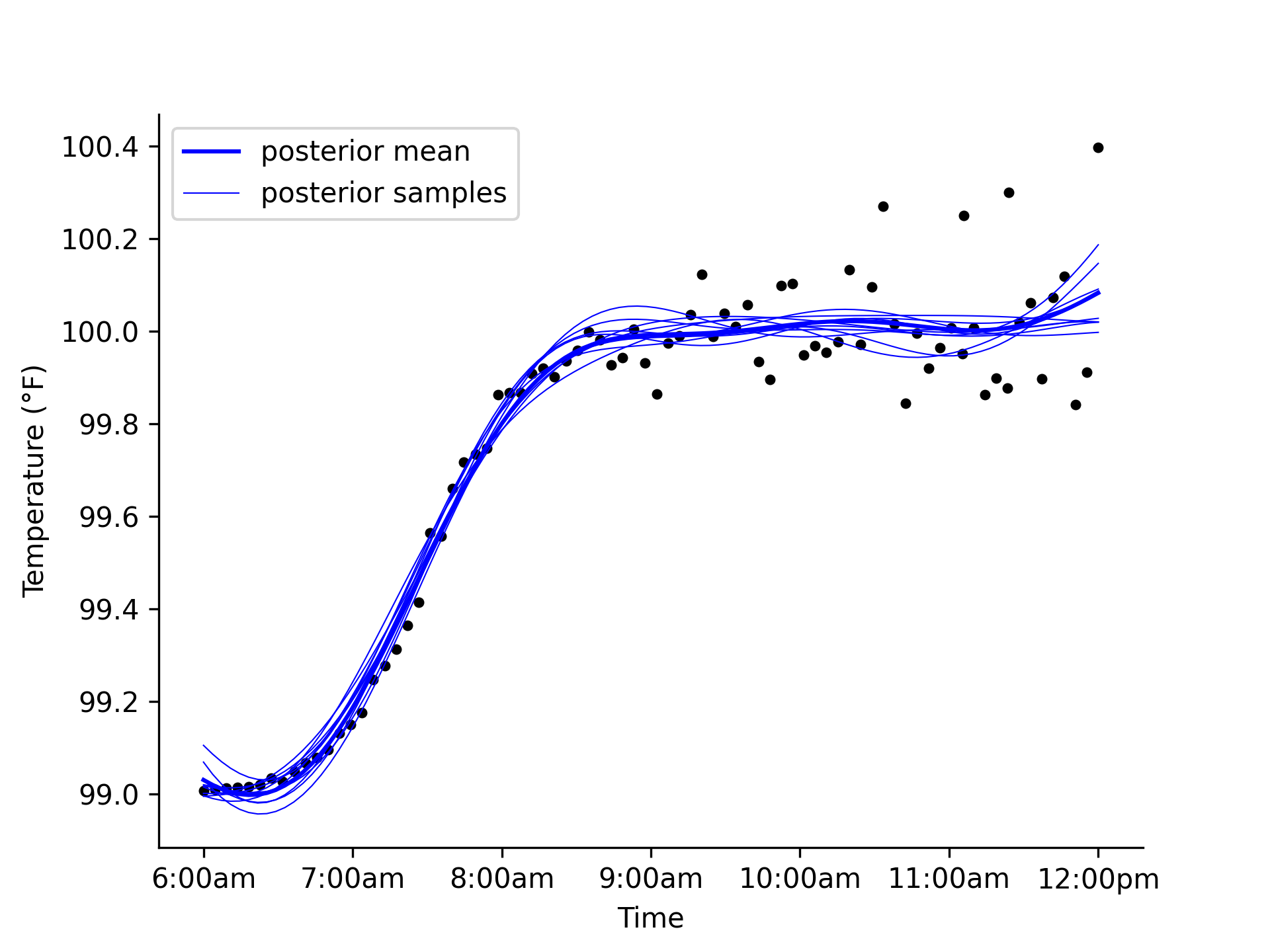}
    \caption{original}
    \label{fig:gp_samples1a}
\end{subfigure}
\hfill
\begin{subfigure}[b]{0.32\textwidth}
    \centering
    \includegraphics[width=\textwidth]{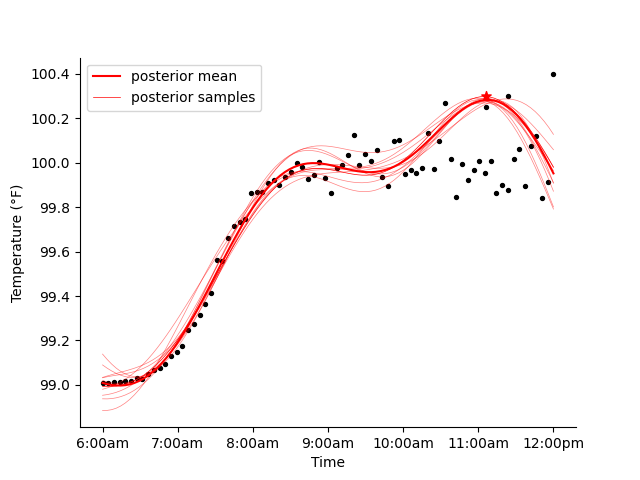}
    \caption{one pseudo-data point}
    \label{fig:gp_samples1b}
\end{subfigure}
\hfill
\begin{subfigure}[b]{0.32\textwidth}
    \centering
    \includegraphics[width=\textwidth]{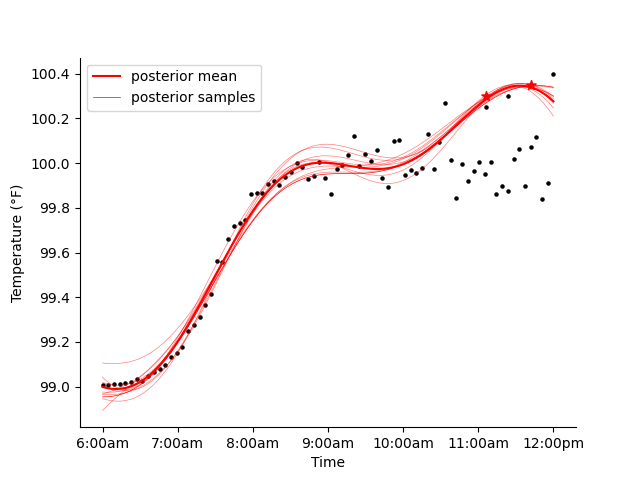}
    \caption{two pseudo-data points}
    \label{fig:gp_samples1c}
\end{subfigure}
\caption{\em Fitted curves corresponding to draws from the GP posterior with pseudo-data, fixing the fitted hyperparameters of the covariance kernel.}
\label{fig:gp_samples1}
\end{figure}

The simulated data is thermometer recordings taken every few minutes for several hours
from one (potentially sick) person. The readings are noisy (see Figure \ref{fig:gp_samples1})
and we use a Gaussian process model to estimate the true temperature. Specifically, we use
the following model:
\begin{align}
y & \sim \text{normal}(f(t), \sigma) \\
f(t) & \sim \mathcal{GP}(98.6, k),
\end{align}
where $k$ is the squared exponential kernel \cite{Rasmussen:2005},
\begin{align}
    k(t, t') = \alpha \exp\left(\frac{-(t - t')^2}{2\ell^2}\right).
\end{align}
%
We fit the hyperparameters of the model---the variance $\alpha$,
the length scale $\ell$, and the residual standard deviation $\sigma$---by maximizing
the log marginal likelihood. That is, assuming $\theta = (\alpha, \ell, \sigma)$, we maximize the function
(defined up to an additive constant)
\begin{align}
\log p(y | \theta) = -\frac{1}{2} y^T (K(\theta) + \sigma^2 I)^{-1} y - \frac{1}{2} \log | K(\theta) + \sigma^2 I |
\end{align}
over $\theta$.

After fitting hyperparameters, we sample from the GP, conditional on the observed data, fitted hyperparameters, and residual standard deviation (see Figure \ref{fig:gp_samples1a}).
The posterior draws look reasonable, other than the large observation noise near 6am and the small observation noise near noon.
%
%
The posterior draws estimate a temperature around $100 \degree$F after around 9am.
However, a possible concern is the upward-sloping temperature recordings above
$100.2 \degree$ F towards the end of the interval.
The posterior doesn't place much mass near these points, but a nervous doctor might want
to get a sense why. Does the Gaussian process prior not allow it? Are there too few recordings
there? How ``easy" would it be to move the posterior through these points?
To get a sense of the answer for these questions, we use the Squealer.

Adding the pseudo-data
point $(x^*, y^*)$, we now maximize a new log marginal likelihood (now marginalizing over $N+1$ instead of $N$ dimensions) where
\begin{align}
y \to
\begin{bmatrix}
y \\
y^*
\end{bmatrix}, \qquad
K(\theta) \to
\begin{bmatrix}
K(\theta) & k(x^*, x)\\
k(x^*, x)^T & k(x^*, x^*)
\end{bmatrix}, \qquad
\sigma^2 I \to
\begin{bmatrix}
\sigma^2 I & 0 \\
0 & 10^{-6}
\end{bmatrix}.
\end{align}
Here, $k(x^*, x)$ is a column vector whose $i^{\text{th}}$ element is $k(x^*, x_i)$. We use
a variance of $10^{-6}$ for the pseudo-data point which has the effect of strongly dragging
the posterior through $(x^*, y^*)$.

In Figure \ref{fig:gp_samples1b}, we add a pseudo-data point at 11am of $100.3\degree$F,
adjusting the posterior the posterior so that it moves through the pseudo-data point in a smooth
way. But the posterior
still doesn't ``want to" go through the points over $100.2\degree$F. The smoothness imposed by the squared-exponential kernel still allows the posterior to return to its former posterior mean (without a pseudo-data point) by noon, the end of the interval.
When we add a second pseudo-data point (see Figure \ref{fig:gp_samples1c}), the posterior
now goes roughly through the points above $100.2\degree$F.

Table \ref{t:hyperparams}
shows the fitted hyperparameter values for the original posterior, and the posteriors with
one and then two pseudo-data points.
The residual standard deviation, $\sigma$, increases from $0.10$ to $0.14$ to $0.17$ when
going from zero to one to two pseudo-data points. So the model is allowing the posterior
to go through the pseudo-data points by fitting a larger observation noise.
In other words, in order to get the posterior to go through the high temperature points,
the pseudo-data points forced the posterior observation noise to be larger than the original
data indicated.
The combination of the normal observation model and a large number of measurements
just less than $100\degree$F makes the original posterior concentrate on a smaller residual
standard deviation and puts little weight on high temperatures.

\begin{table}
\centering
\begin{tabular}{lccc}
 & $\ell$ & $\alpha$ & $\sigma$ \\\hline
original & 0.23 & 0.52 & 0.10 \\
one pseudo-point & 0.23 & 0.48 & 0.14 \\
two pseudo-points & 0.26 & 0.58 & 0.17
\end{tabular}
\caption{\em Fitted values of hyperparameters, timescale $\ell$, variance $\alpha$, and residual standard deviation $\sigma$ for the original model and for both versions with pseudo-data.}
\label{t:hyperparams}
\end{table}

\section{Discussion}

The Squealer uses sonic feedback to supplement rather than replace a visual display. Often it is not immediately clear why a model is not fitting a dataset as the practitioner expects, or why the model is neglecting some fit that seems obvious to the user's eye. Even with very expressive models like Gaussian processes, the factors constraining how they can fit data are opaque, especially for the non-expert. Allowing the modeler to make the curve go where they think it should can sometimes resolve this confusion, for instance by demonstrating that certain other points are very poorly fit by the new curve. But the likelihood may suffer in ways that are not visually apparent. Supplementing the visual fit with auditory feedback can give the modeler helpful intuition as to why the model behaves as it does.

In the present paper we have demonstrated the principles of Squealer with static graphics, but ultimately it is designed to operate dynamically, so that a user can grab, move, and pin a fitted curve, see the consequences for the estimated parameters and the fit to individual data points, and hear the decline in the overall fit.

Several challenges remain to reach that goal.

First, with a tool that's meant to be interactive, the program needs to quickly converge to the updated best fit or posterior distribution quickly to allow smooth operation.  Using existing software this is done by nested loops, with the outer loop being the steps that move the curve as it is being dragged, and the inner loop being the iterations of Stan (or some other probabilistic program) fitting the model at each step along that path.  It would make more sense to thread these loops, so that the target distribution is gradually shifting while the simulations are happening, in the manner of sequential Monte Carlo \parencite{Liu:1998}.

Second, details of the Squealer will depend on the problem under study.  In the same way that maximum likelihood, variational inference, or Bayesian inference are general methods that have different implementations for different models, so must the computation and visual display of the Squealer be adapted to the mathematical structure and goals of the models being fit.  We suspect that the best way to develop the Squealer will be to begin by setting it up for particular classes of models and then ultimately designing a more general implementation.

Third, we would like the different parts of the dashboard to work in concert, so that, in addition to being able to grab and pin fitted curves and see how these shift the fitted curves and change the fit to individual data points, the user can also shift and pin parameter values directly.  Again, it will make the most sense to first implement this for a particular example.

Fourth, as discussed in Section \ref{perturbing.posterior}, the gravitational implementation of the Squealer causes the posterior distribution to be constricted in the range of the pseudo-data points.  We would like to develop a method for shifting the posterior that does not suffer from this artifact.

Finally, the sensification could be done in different ways.  One appealing idea is haptic feedback, where the user would use a joystick to drag the curve, and moving it away from its best fit would induce resistance on the controller so that the user would physically feel the resistance to degrading the fit to prior and data.

Our larger goal is to be able to look under the hood of models better to see why they don't fit as we'd expect. Specifically, we want to see which parameters suffer from the fit we want to impose on the model, and why. This could be a valuable step in the statistical and machine-learning workflow in the modern computing environment in which probabilistic programming tools allow us to fit increasingly complicated models.

\printbibliography

@book{Rasmussen:2005,
    author = {Rasmussen, Carl Edward and Williams, Christopher K. I.},
    title = {Gaussian Processes for Machine Learning},
    publisher = {MIT Press},
    year = {2005}
}

@article{Broadie:2018,
  author = {Broadie, Mark},
  title = {Two simple putting models in golf},
  note = {\url{https://statmodeling.stat.columbia.edu/wp-content/uploads/2019/03/putt_models_20181017.pdf}},
  year = 2018
}

@article{Gelman:2019,
  author = {Gelman, Andrew},
  title = {Model building and expansion for golf putting},
  journal = {Stan Case Studies},
  volume = 6,
  note = {\url{https://mc-stan.org/users/documentation/case-studies/golf.html}},
  year = 2019
}

@book{Workflow:2025,
author={Gelman, Andrew and Vehtari, Aki and McElreath, Richard and others},
year=2026,
title={Bayesian Workflow},
  address = {London},
  publisher = {CRC Press}
}

@article{Vis:2023,
year={2023},
title={From visualization to sensification},
journal={Amstat News},
number={547},
pages={18--19},
author={Andrew Gelman and S. Gwynn Sturdevant}
}

@article{Delivering:2022,
year={2022},
title={Delivering data differently},
author={S. Gwynn Sturdevant and A. Jonathan R. Godfrey and Andrew Gelman},
note={\url{https://arxiv.org/abs/2204.10854}}
}

@article{Dilution:2004,
year={2004},
title={Bayesian analysis of serial dilution assays},
journal={Biometrics},
volume={60},
pages={407--417},
author={Andrew Gelman and Ginger Chew and Michael Shnaidman}
}

@article{Typical:2020,
  author = {Gelman, Andrew},
  title = {The typical set and its relevance to Bayesian computation},
journal={Statistical Modeling, Causal Inference, and Social Science},
  note = {2 Aug.  \url{https://statmodeling.stat.columbia.edu/2020/08/02/the-typical-set-and-its-relevance-to-bayesian-computation/}},
  year = 2020
}

@article{Exploratory:2004,
title={Exploratory data analysis for complex models (with discussion)},
journal={Journal of Computational and Graphical Statistics},
volume={13},
pages={755--787},
author={Andrew Gelman},
year={2004}
}

@article{Wickham:2006,
  author = {Wickham, Hadley},
  title = {Exploratory model analysis with R and GGobi},
  note = {\url{https://had.co.nz/model-vis/2007-jsm.pdf}},
  year = 2006
}

@article{Wickham-Cook-Hofmann:2015,
  author = {Wickham, Hadley and Cook, Dianne and Hofmann, Heike},
  year = 2015,
  title = {Visualizing statistical models: Removing the blindfold},
  journal = {Statistical Analysis and Data Mining},
  volume = 8,
  pages = {203--225}
}

@article{Liu:1998,
title={Sequential Monte Carlo methods for dynamic systems},
author={Liu, Jun S. and Chen, Rong},
journal={Journal of the American Statistical Association},
volume={93},
year={1998},
pages={1032--1044}
}

@book{Dodelson:2003,
  author    = {Dodelson, Scott},
  title     = {Modern Cosmology},
  publisher = {Academic Press},
  address   = {San Diego},
  year      = {2003}
}

@article{HuSugiyamaSilk:1997,
  author  = {Hu, Wayne and Sugiyama, Naoshi and Silk, Joseph},
  title   = {The physics of microwave background anisotropies},
  journal = {Nature},
  volume  = {386},
  pages   = {37--43},
  year    = {1997},
  doi     = {10.1038/386037a0},
  eprint  = {astro-ph/9504057}
}

@article{BJK:2000,
  author  = {Bond, J. R. and Jaffe, A. H. and Knox, L.},
  title   = {Radical compression of cosmic microwave background data},
  journal = {Astrophysical Journal},
  volume  = {533},
  pages   = {19},
  year    = {2000},
  doi     = {10.1086/308625},
  eprint  = {astro-ph/9808264}
}

@article{MASTERCMB,
  author  = {Hivon, E. and G\'orski, K. M. and Netterfield, C. B. and Crill, B. P. and Prunet, S. and Hansen, F.},
  title   = {{MASTER} of the cosmic microwave background anisotropy power spectrum: A fast method for statistical analysis of large and complex cosmic microwave background data sets},
  journal = {Astrophysical Journal},
  volume  = {567},
  pages   = {2},
  year    = {2002},
  doi     = {10.1086/338126},
  eprint  = {astro-ph/0105302}
}

@article{NAMASTER,
  author  = {Alonso, David and Sanchez, Javier and Slosar, An\v{z}e},
  title   = {A unified pseudo-{$C_\ell$} framework},
  journal = {Monthly Notices of the Royal Astronomical Society},
  volume  = {484},
  pages   = {4127--4151},
  year    = {2019},
  doi     = {10.1093/mnras/stz093},
  eprint  = {1809.09603}
}

@article{PlanckCosmology,
  author  = {{Planck Collaboration} and Aghanim, N. and Akrami, Y. and Ashdown, M. and others},
  title   = {{Planck} 2018 results. {VI}. Cosmological parameters},
  journal = {Astronomy \& Astrophysics},
  volume  = {641},
  pages   = {A6},
  year    = {2020},
  doi     = {10.1051/0004-6361/201833910},
  eprint  = {1807.06209}
}

@article{PlanckLikelihood,
  author  = {{Planck Collaboration} and Aghanim, N. and Akrami, Y. and Ashdown, M. and others},
  title   = {{Planck} 2018 results. {V}. {CMB} power spectra and likelihoods},
  journal = {Astronomy \& Astrophysics},
  volume  = {641},
  pages   = {A5},
  year    = {2020},
  doi     = {10.1051/0004-6361/201836386},
  eprint  = {1907.12875}
}

@article{Bornmann2024,
author={Lutz Bornmann},
title={The sound of science},
journal={EMBO Reports},
volume={25},
pages={3743--3747},
year={2024}
}

@article{Daye2006,
author={Christian Dayé and Alberto {de Campo}},
title={Sounds sequential: sonification in the social
sciences},
journal={Interdisciplinary Science Reviews},
year={2006},
volume={31},
pages={349--364}
}

\end{document}